\documentstyle[epsf,epsfig,eqsecnum,aps]{revtex}
\begin{document}
\preprint{\vbox{\hbox{DOE/ER/40762-189}\hbox{UMD PP\#00-007}}}

\title {Looking For Disoriented Chiral Condensates From Pion Distributions}
\author{Chi-Keung Chow and Thomas D.~Cohen}
\address{Department of Physics, University of~Maryland, College~Park, 
MD~20742-4111}
\date{\today}
\maketitle
\begin{abstract}
We suggest two methods for the detection of the formation of disoriented 
chiral condensates in heavy ion collisions.  
We show that the variance in the number of charged pions (in a suitable range 
of momentum space) provides a signature for the observation of a disoriented
chiral condensate.  
The signal should be observable even if multiple domains of D$\chi$C form 
provided the average number of pions per domain is significantly larger than 
unity.  
The variance of the number charged pions alone provides a signal which can be 
used even if the number of neutral pions cannot be measured in a given 
detector.   
On the other hand, the probability distribution in $R$, the proportion of 
neutral pions to all pions emitted in heavy ion collisions in certain 
kinematic regions, has been suggested as a signal of a disoriented chiral 
condensate.  
Here we note that the signature can be greatly enhanced by making suitable 
cuts in the data.  
In particular, we consider reducing the data set such that the $k$ pions with 
lowest $p_T$ are all neutral. 
We find that, given such cuts, $\langle R \rangle$ can be substantially 
different from $1/3$.  
For example, for a single D$\chi$C domain without contamination due to 
incoherently emitted pions, $\langle R \rangle$ is $3/5$ given the pion with 
lowest $p_T$ is neutral, and $5/7$ given the two pions with lowest $p_T$ 
are both neutral, {\it etc.}. 
The effects of multi-domain D$\chi$C formation and noise due to incoherent 
pion emission can be systematically incorporated.  
Potential applications to experiments and their limitations are briefly 
discussed.  
\end{abstract}
\pacs{}
\section{introduction}

During the past several years, there has been considerable excitement about the
possibility of the formation of disoriented chiral condensates (D$\chi$C) in 
heavy ion collisions 
\cite{And,KK,An1,An2,BK,Bj1,Bj2,RW1,KT,RW2,BKT1,GGM,BKT2,GGP,GM,CBNJ,AK,CKMP,AHW,R,C,Bj3}.  
The basic scenario is as follows:  In an ultrarelativistic heavy ion collision 
some region thermalizes at a temperature above the chiral restoration 
temperature.  
If the system cools sufficiently rapidly back through the transition 
temperature, the region will remain in a chiral restored phase.
However, this phase is unstable; small fluctuations in any chiral direction
($\sigma,\vec{\pi}$) will grow exponentially.  
This can create regions where the pion field has macroscopic occupation.
It should be stressed that this scenario is not derivable directly from the
underlying theory of QCD and contains a number of untested dynamical
assumptions, principally that the cooling is rapid.  
Thus the failure of the system to form a D$\chi$C cannot be used to rule out 
that the system has reached the chiral restoration temperature.  
On the other hand,  observation of the formation of a D$\chi$C would be clear 
evidence that the phase transition had been reached.  
  
Unfortunately, since the scenario is not derived directly from a 
well-defined theory, it is difficult to know precisely what constitutes 
observation of a D$\chi$C.
Assuming the system forms a single large domain of D$\chi$C containing a large 
number of pions there should be clear signatures.  
In the first place, one expects an excess in the number of low $p_T$ pions 
produced.
They would be at low $p_T$ since, by hypothesis, the region is large so the 
characteristic momentum is small; the excess would be measured relative 
to a purely statistical thermal distribution.  
Such a signal has the advantage of working even if multiple regions of 
D$\chi$C form provided each region is large enough so the characteristic 
momentum is sufficiently small to provide a discernible signal over the 
thermal background.  
Such a signal is not decisive since one could imagine some other 
collective low energy effects which produce low $p_T$ pions.    

A much stronger signal of a single large domain of D$\chi$C has been proposed. 
Since the pions formed in a D$\chi$C are essentially classical they form a 
coherent state.  
The coherent state has some orientation in isospin space (or more 
precisely the system is a quantum superposition of coherent states with 
different orientations and particular correlations to the isospin of the 
remainder of the system \cite{CBNJ,C}).  
In essence all of the pions in the domain are pointing in the same 
isospin direction.  
Provided the total number of pions in the domain is large, this implies 
that the distribution of the ratio of neutral to total pions in the domain is 
given by \cite{And,KK,An1,An2,Bj1,Bj2,RW1,R} 
\begin{equation}
f(R) \, =  \, \frac{1}{2 \sqrt{R}}
\label{PofR}
\end{equation}
where R is the ratio of the number of $\pi_0$'s in the D$\chi$C divided
 by the total number of pions and $f(R)$ is the probability.
The derivation of $f(R)$ is quite simple and will be discussed below.  
The distribution in Eq.~(\ref{PofR}) is qualitatively distinct from 
a purely statistical distribution in which the emission of charged and neutral 
pions is uncorrelated.  
The distribution from uncorrelated emissions in the infinite particle number 
limit approaches a delta function at $R=1/3$.  
For finite (but large) particle number the statistical distribution is 
narrowly peaked about 1/3 with a variance, $\langle R^2 \rangle - \langle R 
\rangle^2 = \frac{2}{9{\cal N}}$ where $\cal N$ is the total number of pions.  
Since these two distributions are so radically different one should in 
principle have a very clear signal if a single region of D$\chi$C where to 
form in heavy ion reactions and if the pions from the D$\chi$C are 
kinematically separated from other pions in the system.  

The dramatic nature of the preceding signature is based in large measure
on the assumption that a single large domain of D$\chi$C is formed.  
{\it A priori} this seems rather unlikely for the following reason: 
If a large region of the system starts in a hot chirally restored phase and 
then rapidly cools through the phase transition, then there will be a large 
region which is unstable against growth of the pion field.
Presumably, this happens as a ``seed'' fluctuation in a small region which 
rapidly grows.  
It takes a time of at least $L/c$ for information about the formation of the 
domain to propagate a distance $L$.  
However during the time this fluctuation is growing out to $L$, the pion field 
at $L$ has been sitting in an unstable situation.  
The characteristic time it can remain in this unstable configuration is 
$\tau$, the exponential growth time.  If the information about the initial 
seed does not reach $L$ is a time comparable to $\tau$ the region near $L$ 
will likely begin its own exponential growth but in a chiral direction 
uncorrelated from the initial growth.  
Thus, one expects domains of characteristic size $c\tau$ \cite{GGP,GM}.  

The effect of multiple domains on the $R$ distribution is fairly clear: it will
tend to wash out the signal.  
If a large number of domains form and the pions emerging from different 
domains cannot be distinguished kinematically it is clear from the central 
limit theorem that the $R$ distribution will approach a normal distribution.  
This normal distribution may be distinguished from the normal distribution 
arising from uncorrelated emission; the case of multiple domains of D$\chi$C 
will have a substantially larger variance.

Unfortunately, there is an important practical limitation which makes it 
difficult to exploit the $R$ distribution as a signature.   
Even under the most optimistic of scenarios, the total number of pions coming 
from D$\chi$C's will be a small fraction of the total number of pions.  
If one includes all pions produced in the reaction, the signal from the pions 
from the D$\chi$C will presumably be overwhelmed.  
Thus, it is highly desirable to use kinematic consideration to enhance the 
contributions coming from the D$\chi$C.  
In particular, it is sensible to study the $R$ distribution for a sample 
restricted to low $p_T$ pions only.
In any scenario where the D$\chi$C is well defined, {\it i.e.}, the occupation 
number is large is likely to require a moderately large regions of D$\chi$C 
and the characteristic momentum spread in the D$\chi$C will be fixed by the 
inverse size of the region.  
Thus one expects D$\chi$Cs to preferentially produce moderately low $p_T$ 
pions.  
(One also should restrict the pions in the distribution to a moderately 
narrow rapidity window). 

As an experimental matter, it should be relatively straightforward to cut on 
the momentum of the charged pions to select low $p_T$ pions in a given 
rapidity window.  
For neutral pions, however, it is not a simple matter.  
The neutral pions will decay in flight and will ultimately be detected as 
photons.  
If one is simply interested in the overall $R$ distribution, without cuts, 
and if the detected photons come predominately from $\pi^0$ decays then one 
can use the $n_\gamma/2$ as a surrogate for $n_{\pi^0}$.  
Recent experimental searches have exploited this strategy\cite{MM,Bj4,WA}.  
However, in order to study the $R$ distribution in a limited kinematical 
region it is necessary to reconstruct the $\pi^0$ momenta from the observed 
photons in order to make kinematical cuts on the $\pi^0$ momenta.  
Since the number of neutral pions per event is large, the reconstruction of 
neutral pion momenta is likely to be a formidable task.
This raises the following interesting question:  
{\sl Can one find a signature for the presence of regions of D$\chi$C of 
essentially the same quality as the $R$ distribution but which does not 
require the measurement of neutral pions?}

On the other hand, even if one succeeds in reconstructing the neutral pions 
so that a low $p_T$ cut can be applied, the noise may still be severe, 
as both the signal and the noise peak at $R=1/3$.  
So it is highly desirable to have some other cuts to further suppress the 
noise due to incoherently emitted pions. 
This raises another question: {\sl Can one construct additional cuts to 
further enhance the signal-to-noise ratio?}  

In this paper, we will provide answers to both questions raised above.  
For the first question, we will show in Sec.~II that 
the distribution of the number of charged pions (in a kinematically limited 
region) contains essentially the same information about D$\chi$C formation as 
the $R$ distribution.  
This should greatly aid in searches for D$\chi$C formation.
We also discuss additional information that can be inferred if $\pi^0$'s can
be reconstructed.  
In particular, we show that the distribution of the total number of pions (in 
a limited kinematic region) provides a means to distinguish D$\chi$C formation 
from other hypothetical mechanisms for the production of a pion coherent 
state.
Then in Sec.~III, we will suggest cuts which may dramatically enhance the 
signal-to-noise ratio.  
We study the conditional probability distribution of $R$ given only for the 
events in which the $k$ pions with the lowest $p_T$ are all neutral, and we 
will show that the expectation value of $R$ is shifted away from $1/3$.  
Since incoherent emission will result in a very narrow peak around $R=1/3$, 
any such shifts should be easily observable.  
Moreover, one can make successive cuts by increasing the value of $k$, and 
enhance the signal in each successive step. 
We believe these signatures will be useful in the searches for D$\chi$C 
formation in RHIC and LHC.  

\section{Signature of Disoriented Chiral Condensates From Charged Pions}

In this section, we study the distribution of the number of charged 
pions produced from a D$\chi$C.  
We show that it contains essentially the same information about D$\chi$C 
formation as the distribution of $R$, and the distribution will be much wider 
in the presence of D$\chi$C formation than otherwise. 
This is of significance since it provides a signature which avoids the 
experimentally challenging task of reconstructing neutral pions. 

\subsection{The Simple Case of a Single Domain}

We begin by studying an overly simple model and in subsequent subsections we 
will generalize our results to more realistic scenarios.  
In this simplified situation we assume that in every collision a single large 
domain of D$\chi$C is formed with a large particle number.  
Moreover, we will assume that the field strength and spatial distribution of 
this domain do not vary event by event, and that the pions produced in the 
D$\chi$C are kinematically completely distinguishable from all other pions in 
the system (including those pions produced from ``$\sigma$'s'' --- {\it i.e.}, 
fluctuations in the $\langle \overline{q} q \rangle$ directions).  
Finally, we will assume that both isospin violating effects and explicit 
chiral symmetry breaking are negligible.  

By hypothesis, the region of D$\chi$C contains many particles and is
essentially classical in nature. 
To simplify discussion we will adopt the usual convention of 
describing the physics in terms of the degrees of freedom in a linear sigma 
model with O(4) symmetry, {\it i.e.}, $\sigma$ and $\vec{\pi}$ 
rather than directly in terms of the QCD degrees of freedom. 
We wish to stress, however, that we are not relying on the detailed dynamics 
of any particular variant of the $\sigma$ model.

Since a D$\chi$C has macroscopic pion occupation numbers, it is tempting 
to describe it as a coherent state in the O(4) chiral space as in 
Ref.~\cite{GGM}.  
However, in a generic heavy ion collision, the D$\chi$C (if produced) in the 
interior of the fireball will be quantum mechanically correlated with the 
high momentum emission at the edge of the fireball \cite{CBNJ}.  
As a result, the low $p_T$ part of the physical state is not a pure state 
and is more appropriately described by a density matrix $\rho$.  
For example, when explicit chiral symmetry breaking due to non-zero quark 
masses are ignored, a chirally restored state would be described by the 
density matrix $\rho_{\rm sym}$ such that 
\begin{equation}
\langle \sigma \rangle = {\rm Tr}(\rho_{\rm sym}\sigma) = 0, \qquad
\langle \vec\pi \rangle = {\rm Tr}(\rho_{\rm sym}\vec\pi) = 0;  
\end{equation}
and for the physical vacuum, 
\begin{equation}
\langle \sigma \rangle = {\rm Tr}(\rho_{\rm vac}\sigma) = f, \qquad
\langle \vec\pi \rangle = {\rm Tr}(\rho_{\rm vac}\vec\pi) = 0,  
\end{equation}
where the non-zero expectation value of $\sigma$ reflects that chiral 
symmetry is spontaneously broken.  

Now we are ready to study D$\chi$C states $\rho_{D\chi C}
(\psi,\theta,\phi)$, where the three angles describe the orientations of the 
D$\chi$Cs in the O(4) chiral space.  
With the notation $\langle X \rangle (\psi,\theta,\phi) = 
{\rm Tr}(\rho_{D\chi C}(\psi,\theta,\phi)X)$, where $X$ is any operator, 
a D$\chi$C state is characterized by the following relations.  
\begin{eqnarray}
\langle \sigma \rangle (\psi,\theta,\phi)&=& F \cos\psi, \nonumber\\
\langle \pi_0 \rangle (\psi,\theta,\phi) &=& F \sin\psi \cos\theta, \nonumber\\
\langle \pi_x \rangle (\psi,\theta,\phi) &=& F \sin\psi \sin\theta \cos\phi, \\
\langle \pi_y \rangle (\psi,\theta,\phi) &=& F \sin\psi \sin\theta \sin\phi, 
\nonumber
\end{eqnarray}
where $F^2 = \langle \sigma \rangle^2 + \langle \pi_0 \rangle^2 
+ \langle \pi_x \rangle^2 + \langle \pi_y \rangle^2$ is independent of the 
angles $\psi$, $\theta$ and $\phi$ but can depend on space and time.  

The number operators for neutral and charged pions are 
\begin{equation}
n_0 \sim \pi_0^\dag \pi_0, \qquad 
n_\pm \sim \pi_x^\dag \pi_x + \pi_y^\dag \pi_y, 
\end{equation}
and one can easily find their expectation values , {\it i.e.}, $\langle 
n_{0,\pm} \rangle(\psi,\theta,\phi) = {\rm Tr}(\rho_{D\chi C}
(\psi,\theta,\phi) n_{0,\pm})$.
It turns out that $\langle n_{0,\pm}\rangle$ can be factorized into the 
following form.  
\begin{equation}
\langle n_{0,\pm} \rangle(\psi,\theta,\phi) = \langle n \rangle \, 
g_{0,\pm}(\psi,\theta,\phi), 
\label{fac}
\end{equation}
where 
\begin{equation}
\langle n \rangle = \langle n_0 \rangle (\psi=\pi/2,\theta=0,\phi) 
= \langle n_\pm \rangle (\psi=\pi/2,\theta=\pi/2,\phi)
\end{equation}
and $\phi$ can take on arbitrary values.  
In other words, the expectation value $\langle n \rangle$ measures the total 
number of $\pi$'s produced by the D$\chi$C if fully oriented in a pionic 
direction.  
In general, $\langle n \rangle$ depends on the dynamical details of the 
D$\chi$C, such as its probability distribution in position or momentum spaces. 
The geometrical factors $g_0(\psi,\theta,\phi)$ and $g_\pm(\psi,\theta,\phi)$ 
will be called the neutral and charged proportions respectively, and they take 
the following forms.  
\begin{equation}
g_0(\psi,\theta,\phi) = \sin^2 \psi \cos^2 \theta, \qquad 
g_\pm(\psi,\theta,\phi) = \sin^2 \psi \sin^2 \theta.
\label{defg}
\end{equation}
Note that $\langle n \rangle$ does not depend on the 
orientation angles $(\psi, \theta, \phi)$ while $g_{0,\pm}$ do not depend on 
the dynamical details of the D$\chi$C.  

Equation (\ref{fac}) represents that both expectation values $\langle n_0 
\rangle$ and $\langle n_\pm \rangle$ are related to $\langle n \rangle$, 
and hence to each other, up to geometrical factors $g_{0,\pm}$.  
Actually the relationship goes deeper than that: the probability 
distributions of the number of neutral and charged pion produced, denoted by 
$p_0(n_0)$ and $p_\pm(n_\pm)$ respectively, can both be expressed in terms 
of the same distribution $p(n)$.  
\begin{equation}
p_{0,\pm}(n_{0,\pm};\psi,\theta,\phi) = 
p(n_{0,\pm}/g_{0,\pm}(\psi,\theta,\phi)) \big/ g_{0,\pm}(\psi,\theta,\phi), 
\quad \int dn_{0,\pm} p_{0,\pm}(n_{0,\pm};\psi,\theta,\phi) = \int dn p(n) = 1.
\end{equation}
Notice that $p(n)$ does not depend on the angles, and all the angular 
dependences in $p_{0,\pm}$ come from the geometrical factors $g_{0,\pm}$.  
One can easily see that the above relation implies Eq.~(\ref{fac}), and 
also the following relation on the higher moments: 
\begin{equation}
\langle n_{0,\pm}^k \rangle(\psi,\theta,\phi) = \langle n^k\rangle \,
g^k_{0,\pm}(\psi,\theta,\phi), 
\end{equation}
with 
\begin{equation}
\langle n^k \rangle = \int n^k dn p(n) .  
\end{equation}

The probability distribution $p(n)$ determines the distributions of neutral 
and charged pions produced from a D$\chi$C with a given orientation (as 
described by the angles $(\psi,\theta,\phi)$) in the O(4) chiral space.  
Let's recall that the probability distributions of the numbers of neutral 
or charged pion incoherently emitted in a heavy ion collision are 
Poisson--Gaussian, which have narrow peaks.  
To quantify the narrowness of a distribution, it is useful to define the 
{\it deviance\/} $\delta[X]$ of a distribution of variable $X$ such that 
\begin{equation}
\langle X^2 \rangle = (1+\delta[X]) \langle X \rangle^2, \qquad \hbox{or} 
\qquad (\Delta X)^2 \equiv \langle X^2 \rangle - \langle X \rangle^2 
= \delta[X] \langle X \rangle^2.   
\end{equation}
For a Poisson--Gaussian distribution, 
\begin{equation}
\langle n^2\rangle = \langle n \rangle^2 + \langle n \rangle, \qquad
\delta[n] = 1/\langle n \rangle \to 0 \quad \hbox{when $\langle n \rangle \to 
\infty$}. 
\end{equation}
If the distribution $p(n)$ is wide ({\it i.e.}, with non-zero deviance in the 
large $\langle n \rangle$ limit), the fluctuations in the numbers of 
coherently emitted neutral or charged pions will be much larger than the case 
of incoherent emission.  
As a result, substantial deviances in the number of neutral or charged pions 
produced may signify D$\chi$C formation.  
Unfortunately, the form of $p(n)$ depends on the dynamical details of the 
D$\chi$C and is extremely model-dependent.  
Various forms of coherent states and density matrices have been motivated by 
different theoretical considerations \cite{AK} and lead to various forms of 
$p(n)$.  
In particular, the well-studied Glauber coherent state \cite{G} also give a 
Poisson--Gaussian $p(n)$ and hence negligible deviances in the pion numbers 
in the limit $\langle n \rangle \to \infty$.  
So it may seem impossible to determine whether D$\chi$Cs are formed by 
studying the fluctuations in the number of neutral or charged pions produced.  
However, recall that the above discussion presumes we have a D$\chi$C with 
a definite orientation in chiral space, {\it i.e.}, definite values of $\psi$,
$\theta$ and $\phi$.  
In D$\chi$C formation, however, the orientation is randomly generated, and 
the relevant object of study is an density matrix with some appropriate 
probability distribution functions for the angles, which we will now proceed 
to study.  

\bigskip

The above analysis shows that, {\it for each set of orientation angles 
$(\psi, \theta, \phi)$,} the distributions of $n_{0,\pm}$ are normal.  
However, since the orientation is randomly generated in the process of 
spontaneous symmetry breaking, one cannot predict $(\psi, \theta, \phi)$.  
On the other hand, since we are neglecting explicit chiral symmetry breaking, 
the system is equally probable to point in any direction in chiral space.  
Moreover, since by hypothesis we are in a semiclassical situation (large 
$\langle n \rangle$), it is legitimate to work with probabilities rather 
than amplitudes.  
Using the technology of Ref.~\cite{R}, the probability 
distribution in the angular variables is given by the unit measure:  
\begin{equation}
d^3P(\psi,\theta,\phi) = \frac{1}{2 \pi^2} \; \sin^2\psi \, \sin\theta \; 
d\psi \, d\theta \, d\phi , 
\label{pa}
\end{equation}
and the relevant density matrix is 
\begin{equation}
\rho_{D\chi C} = \int d^3P(\psi,\theta,\phi) 
\rho_{D\chi C}(\psi,\theta,\phi).  
\label{domain}
\end{equation} 
One can use Eq.~(\ref{defg}) to 
reparametrize the probability distribution (\ref{pa}) in terms of the 
neutral and charged proportions $g_{0,\pm}=\langle n_{0,\pm}\rangle/
\langle n \rangle$, {\it i.e.}, the fraction of neutral or charged pions among 
all particles produced by the D$\chi$C.  
\begin{equation}
d^2P(g_0,g_\pm) = {1\over \pi} \, {1\over \sqrt{g_0(1-g_0-g_\pm)}} 
\;dg_0\, dg_\pm.
\label{j}
\end{equation}
From this one can obtain the marginal probability distribution of $g_0$ and 
$g_\pm$ by integrating over the other variable.  
\begin{eqnarray}
dP(g_0) & = f_0(g_0) dg_0 = & {2\over \pi} \, \sqrt{1-g_0\over g_0} dg_0, 
\nonumber\\
dP(g_\pm) & = f_\pm(g_\pm) dg_\pm = & dg_\pm.  
\end{eqnarray}
These distribution functions are plotted in Fig.~1 in solid curves.  
It is obvious the both distributions are far from being normal.  
The function $f_0$ is heavily skewed towards the low end and actually 
diverges as $1/\sqrt{g_0}$ when $g_0\to 0$.   
On the other hand, $f_\pm$ is flat, and $g_\pm$ is equally likely to assume 
any value between 0 and $1$.  
This is drastically different from pion emission from an uncorrelated source
(the dotted curves in Fig.~1 \footnote{
For uncorrelated emissions, the probability distributions are Poisson-Gaussian 
with mean $1/4$ and $1/2$ for neutral and charged pions respectively.  
The variances depend on the number of independently emitted pions; the 
plots correspond to the case of $n=50$.}), 
where both distributions would be normal.  

\bigskip
\begin{figure}
\epsfig{file=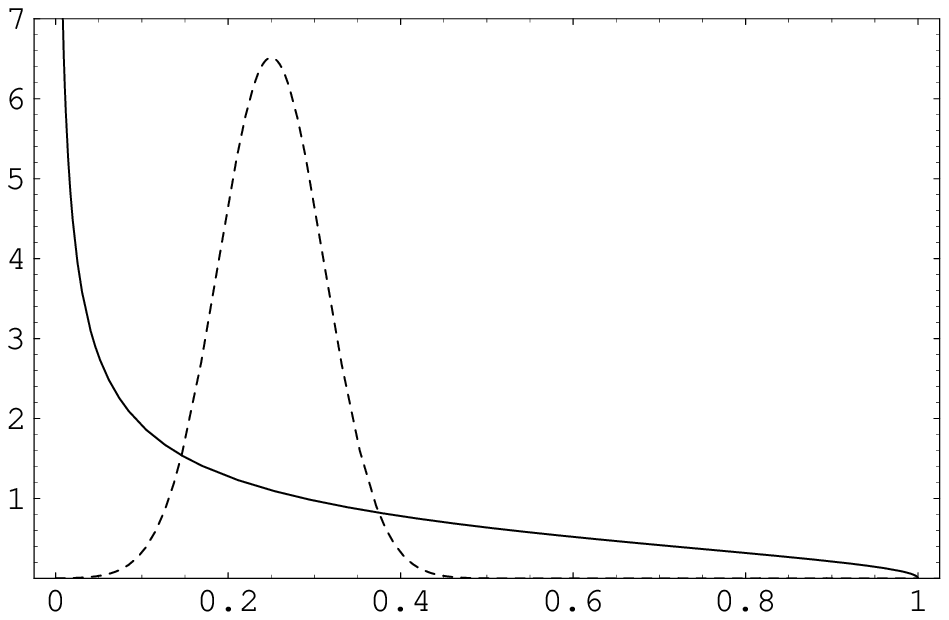, width=3.4in} 
\epsfig{file=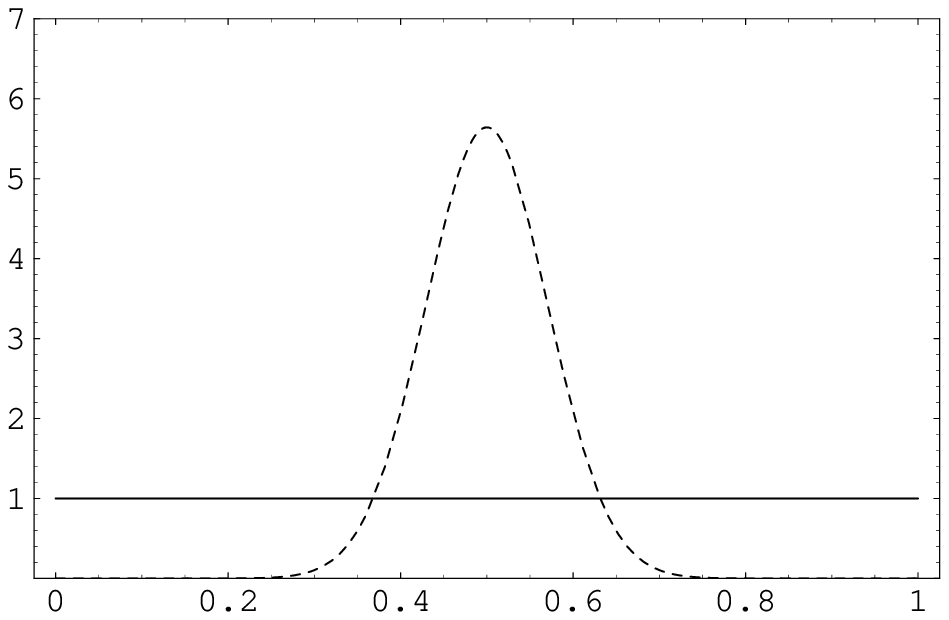, width=3.4in}

\hskip 112pt Fig.~1a \hskip 215pt Fig.~1b

\bigskip

\caption{The probability distribution functions $f_{0,\pm} (g_{0,\pm})$.  
Fig.~1a is $f_0(g_0)$, and Fig.~1b is $f_\pm(g_\pm)$.  
The solid curves are for D$\chi$C emission, while the dotted curves are for 
independent emission with $n=50$.  }
\end{figure}
\bigskip

More quantitatively, one can calculate the first and second moments of 
$g_0$ and $g_\pm$.  
\begin{eqnarray}
\langle g_0 \rangle &= 1/4, \qquad\qquad \langle g_0^2 \rangle =& 1/8, 
\nonumber\\
\langle g_\pm \rangle &= 1/2, \qquad\qquad \langle g_\pm^2 \rangle =& 1/3,   
\end{eqnarray}
and hence the respective deviance $\delta[g_0]$ and $\delta[g_\pm]$, 
\begin{equation}
\delta[g_0] = 1, \qquad \delta[g_\pm] = 1/3.  
\label{d}
\end{equation} 

What does the distributions of the proportions $g_{0,\pm}$ tell us about the 
distributions of $n_{0,\pm}$?  
It is obvious that when $n$ is fixed, $g_{0,\pm}$ give the pion distribution. 
In reality, of course, $n$ is not fixed; we have shown that it behaves like a 
Poisson distribution if we have a conventional Glauber coherent state 
\cite{G}.  
However, since Poisson distributions are sharply peaked, the dispersion of 
$n$ will simply smear the distribution of $n_{0,\pi}$ slightly, without 
changing the overall shape qualitatively.  
More specifically, note that \footnote{
The two set of brackets in the right-hand side of the equations below have 
different physical origins.  
The expectations of $g_{0,\pm}$ are statistical in nature, while that of 
$n$ is via quantum mechanical smearing of a coherent state.  
Consequently the distributions of $n$ and $g_{0,\pm}$ are assumed to be 
uncorrelated, and we derived the equalities below.}  
\begin{eqnarray}
\langle n_{0,\pm} \rangle &=& \langle n \rangle \, 
\langle g_{0,\pm} \rangle, \nonumber\\
\langle n_{0,\pm}^2 \rangle &=& \langle n^2 \rangle \, 
\langle g_{0,\pm}^2 \rangle = (1+\delta[n])\, (1+\delta[g_{0,\pm}]) \; 
(\langle n \rangle \, \langle g_{0,\pm} \rangle)^2 \equiv (1+\delta[n_{0,\pm}])
\langle n_{0,\pm} \rangle^2.  
\end{eqnarray}
The first equality gives 
\begin{equation}
\langle n_0 \rangle = \langle n \rangle / 4, \qquad 
\langle n_\pm \rangle = \langle n \rangle / 2,  
\end{equation}
which have the simple interpretation that, given the symmetry between the 
four directions $(\pi_x, \pi_y, \pi_0, \sigma)$ in the chiral space, 
a quarter of the particles produced by the D$\chi$C will be $\pi_0$, while 
half of them will be $\pi_x$ or $\pi_y$.  
On the other hand, the second equality gives 
\begin{equation}
\delta[n_{0,\pm}] = \delta[n] + \delta[g_{0,\pm}] + \delta[n]\delta[g_{0,\pm}]
\geq \delta[g_{0,\pm}].
\end{equation}
For a Glauber state, $\delta[n] = 1/\langle n\rangle$, $\delta[g_0]=1$ and 
$\delta[g_\pm]$, one has 
\begin{equation}
\delta[n_0] = 1 + 2/\langle n \rangle, \qquad 
\delta[n_\pm] = 1/3 + 4/3\langle n \rangle.  
\label{ds}
\end{equation}
As expected, when $\langle n \rangle$ is large, the deviances 
$\delta[n_{0,\pm}]$ approach $\delta[g_{0,\pm}]$.  
For other states with non-zero $\delta[n]$, the deviances $\delta[n_{0,\pm}]$ 
would be even larger.  
Note that both $\delta[n_{0,\pm}]$ are of order 1, in contrast to an 
uncorrelated emission which would have $\delta = 1/\langle n \rangle$.    
In an ideal world, such enhancements of $\delta$'s would indicate the 
existence of D$\chi$C.  

It is interesting to compare our analysis with that in Ref.~\cite{GGM}, where 
$\delta[n_{0,\pm}]$ (called $C_{00,++}$ in the reference) has also been 
calculated with $\delta[n_0]=4/5$ and $\delta[n_\pm]=1/5$, in contrast to 
our results in Eq.~(\ref{d}) with the values 1 and $1/3$ respectively.  
The discrepancy originates from a key difference in the physical scenarios: 
the states in Ref.~\cite{GGM} is equally likely to point at any direction 
in the O(3) {\it isospace\/} (without a $\sigma$ direction) while our state 
is equally likely to point at any direction in the O(4) {\it chiral space}.  
The extra uncertainty of the orientation of the D$\chi$C in the $\sigma$ 
direction is reflected in the extra angle $\psi$ in addition to the usual 
polar and azimuthal angles $\theta$ and $\phi$ of the sphere in isospace.  
Even with a fixed direction in isospace and hence fixed $\theta$ and $\phi$, 
varying $\psi$ from 0 to $\pi/2$ will change the pion number of both species 
from zero to its maximum value.  
This extra uncertainty in the $\sigma$ direction increases the fluctuation 
in the number of pion produced, leading to larger $\delta[n_{0,\pm}]$ in our 
analysis than their counterparts in Ref.~\cite{GGM}.  
By hypothesis, all direction in the O(4) chiral space {\it with the $\sigma$ 
direction\/} are equally likely, and we believe our formalism is more 
appropriate than that in Ref.~\cite{GGM} in describing a D$\chi$C.  

\subsection{More Realistic Scenarios: Multidomain Formation}

We have studied the simple case of pion emission from a single large (in 
the sense that $\langle n \rangle$ is large) D$\chi$C domain.  
As discussed in the introduction, this scenario is presumably not realistic.  
We will now proceed to study more realistic scenarios with multi-domain 
formation.  
The main point is, even though the probability distribution is smeared out 
because of the lack of alignment (in the chiral space) between the different 
domains, one feature survives, namely the large variance of the distributions.
In particular, we will see that the variances for both neutral and charged 
productions are still much larger than that of an uncorrelated emission.  

Let's consider the case which we have $N$ domains, each with the same 
$\langle n \rangle$.  
We are also assuming both $\langle n \rangle$ and $N$ are much larger 
than unity.  
The total number of pions of each species produced is the sum of pions 
of that particular species produced in each domain, the distribution of 
which has been discussed in the previous section.  
\begin{equation}
\Sigma n_0 = \sum_{i=1}^N n_0^{(i)},\qquad 
\Sigma n_\pm = \sum_{i=1}^N n_\pm^{(i)}.  
\end{equation}
By the central limit theorem, the probability distribution of $\Sigma 
n_{0,\pm}$ will approach normal distributions when $N$ is large.  
However, we will see that the variances of the Gaussian distributions will be 
much larger for pion production from a D$\chi$C than those of uncorrelated 
pion emission. 

Since the pion production in each domain are independent, $n_{0,\pm}^{(i)}$ 
are independent random variables.  
Hence the mean of $\Sigma n_{0,\pm}$ is just the sum of the means of all  
$n_{0,\pm}^{(i)}$, 
\begin{equation}
\langle \Sigma n_{0,\pm} \rangle = \sum_{i=1}^N \langle n^{(i)}_{0,\pm} 
\rangle = N \langle n_{0,\pm} \rangle \equiv {\cal N}_{0,\pm}, 
\label{mean}
\end{equation}
and the variance of the sum $n_{0,\pm}$ is just the sum of the variances of 
each $n_{0,\pm}^{(i)}$.  
\begin{eqnarray}
(\Delta \Sigma n_{0,\pm})^2 &=& \sum_{i=1}^N (\Delta n^{(i)}_{0,\pm})^2 = 
N (\Delta n_{0,\pm})^2 \nonumber\\ &=& N \langle n_{0,\pm} \rangle^2 
\delta [n_{0,\pm}] = (\delta[n_{0,\pm}] / N) {\cal N}_{0,\pm}^2.    
\label{var}
\end{eqnarray}
In other words, 
\begin{equation}
\delta[\Sigma n_{0,\pm}] = {(\Delta \Sigma n_{0,\pm})^2 \over 
\langle \Sigma n_{0,\pm} \rangle^2} = {\delta[n_{0,\pm}] \over N} = 
{\delta[n_{0,\pm}] \langle n_{0,\pm} \rangle \over {\cal N}_{0,\pm}}.  
\label{wide}
\end{equation}
In comparison with uncorrelated pion production, with $\delta = 
1/{\cal N}_{0,\pm}$ we see that for multi-domain D$\chi$C the deviances are 
enhanced by a factor of $\epsilon_{0,\pm} = \delta[n_{0,\pm}] 
\langle n_{0,\pm} \rangle$.  
\begin{equation}
\epsilon_0 = (\langle n \rangle + 2)/4, \qquad 
\epsilon_\pm = (\langle n \rangle + 4)/6.  
\label{e1}
\end{equation}
{\it A priori\/} $\langle n \rangle$ can take any value, but 
$\epsilon_{0,\pm}$ are larger than unity for any value of $\langle n \rangle 
> 2$.  
Even for a very modest $\langle n \rangle =8 $, $\epsilon_0 = 2.5$ and 
$\epsilon_\pm = 2$, leading to substantial widening of the corresponding 
distributions, an observable signature of D$\chi$C formation.  
For larger values of $\langle n \rangle$, the broadening will be even more 
pronounced.  

\bigskip

While the above scenario describes D$\chi$C with multi-domains, a probable 
feature of D$\chi$C formation in the real world (if it happens at all), it is 
still unrealistic in assuming all the domains are of equal strength, 
{\it i.e.}, with the same $\langle n \rangle$.  
Instead one expects $\langle n \rangle$ of different domains to fall under 
a certain probability distribution, which depends on the details of the 
model.  
Naturally, one questions if the signatures discussed above still survive 
under such circumstances.  

Let's consider the case with $N$ domains, with different $\langle n^{(i)} 
\rangle \gg 1$.  
Equation (\ref{mean}) becomes 
\begin{equation}
\langle \Sigma n_{0,\pm}\rangle = \sum_{i=1}^N \langle n^{(i)}_{0,\pm}\rangle
= N \overline {\langle n_{0,\pm} \rangle} \equiv {\cal N}_{0,\pm}, 
\end{equation}
where $\overline {\langle n_{0,\pm} \rangle}$ is the average of  
$\langle n^{(i)}_{0,\pm}\rangle$.  
Equation (\ref{var}) becomes   
\begin{equation}
(\Delta \Sigma n_{0,\pm})^2 = \sum_{i=1}^N (\Delta n^{(i)}_{0,\pm})^2 = 
\sum_{i=1}^N \langle n^{(i)}_{0,\pm}\rangle^2 \delta[n^{(i)}_{0,\pm}].  
\end{equation}
By the inequalities $\delta[n^{(i)}_{0,\pm}] > \delta[g_{0,\pm}]$ 
(cf.~Eq.(\ref{ds})) and $\sum_1^N \langle n \rangle^2 \geq 
(\sum_1^N \langle n \rangle)^2/N$ (mean of squares is larger than square of 
mean), we have 
\begin{equation}
(\Delta \Sigma n_{0,\pm})^2 > \left( \sum_{i=1}^N \langle n^{(i)}_{0,\pm}
\rangle^2 \right) \delta[g_{0,\pm}] \geq \left( \sum_{i=1}^N \langle 
n^{(i)}_{0,\pm}\rangle \right)^2 \delta[g_{0,\pm}] / N = 
(\delta[g_{0,\pm}]/N) {\cal N}_{0,\pm}^2.  
\end{equation}
In other words, 
\begin{equation}
\delta[\Sigma n_{0,\pm}] = {(\Delta \Sigma n_{0,\pm})^2 \over 
\langle \Sigma n_{0,\pm}\rangle^2} \geq {\delta[g_{0,\pm}] \over N} 
= {\delta[g_{0,\pm}] \overline {\langle n_{0,\pm} \rangle} \over 
{\cal N}_{0,\pm}}.  
\end{equation}
(Compare Eq.~(\ref{wide}).)  
Again, the deviances are much larger than that of uncorrelated emission with 
$\delta = 1/{\cal N}_{0,\pm}$ when $\overline {\langle n \rangle} 
\gg 1$ by the following enhancement factors. 
\begin{equation} 
\epsilon_0 \geq \overline {\langle n \rangle}/4, \qquad 
\epsilon_\pm \geq \overline {\langle n \rangle}/6.  
\label{e2}
\end{equation}
So we see that, even with domains of unequal strengths, the number of 
neutral or charged pions produced by a D$\chi$C will still have a much wider 
distribution than that from independent, uncorrelated emission.  

Lastly, one may also ask if the distribution of $\Sigma n_{0,\pm}$ will 
approach normal distributions when $N$ is large in the case of domains 
with unequal strengths.  
In this case $n^{(i)}$'s do not all fall under the same probability 
distribution and the most simple form of the central limit theorem does 
not apply.  
On the other hand, there are generalized forms of the central limit theorem,   
which state that as long as the probability distributions are sufficiently 
``well behaved'', the sum of $N$ random variables will still fall under a 
normal distribution when $N \to \infty$.  
It is actually possible to argue that the distribution of $\Sigma n_\pm$ 
does approach a normal distribution by the Lindeberg generalization of 
the central limit theorem.  
(See, for example, Sec.~6.E of Ref.~\cite{F}.)  
Whether the same conclusion holds for $\Sigma n_0$ is still an open question.  

\bigskip

Let us recapitulate what we have shown:   
one can calculate the probability distribution of the 
number of neutral or charged pions produced as a result of D$\chi$C formation. 
The resultant deviances $\delta$'s are not of the order $1/\cal N$ as in an 
uncorrelated emission, but are instead enhanced by factors $\epsilon$'s 
which are of order $\langle n \rangle$.  
Seeing such enhancements of deviances would be signatures of coherent pion 
productions.  

One can understand the origin of such enhancements of statistical 
fluctuations of the number of neutral or charged pions by considering the 
following analogy.  
Consider two groups of gamblers playing roulette in a casino: $N$ lawyers at 
\$100 tables, and $100 N$ physicists at the \$1 tables, where the odds are 
the same.  
If each lawyer and physicist is given $n$ chips, of \$100 and \$1,  
respectively (so that the total amount given to the lawyers, ${\cal N} = N 
\times 100n$, is the same as that given to the physicists, ${\cal N} = 100N 
\times n$), and is required to bet all of them, the average loss will be the 
same for both groups as long as they are following the same betting 
strategies.  
However, it is easy to see that the statistical fluctuation of the loss 
of the lawyers would be much larger for that of the physicists.  
In other words, the amount of loss, as well as its standard deviation, 
is ``quantized'' in units of the value of the bets.  
The larger the bet, the larger the fluctuation. 
On the other hand, one can also turn the argument around; a discerning 
external observer can deduce, with the knowledge of the gambling strategies, 
the size of the bets of the lawyers from the statistical fluctuations of the 
lawyers' losses, and do likewise for the physicists as well.  

Just as the chips in a single bet of a lawyer share the same fate (either win 
or lose), all the pions in a single D$\chi$C domain share the same orientation 
in the chiral space. 
As a result, the fluctuation of the number of pions in each direction in 
the chiral space is enhanced by a factor proportonal to $\langle n \rangle$, 
the number of pions in each D$\chi$C domain.  
And by reverse argument, one can deduce whether coherent pion emission is 
taking place by measuring the fluctuation of the number of emitted pions.  

\subsection{Including Incoherent Contributions} 

Our analysis has assumed that all the pions originate from coherent emissions, 
and each of these coherent states has $\langle n^{(i)} \rangle \gg 1$.  
In the real world, there is contamination from independent pion emissions, 
and the total numbers of neutral or charged pions are the sums of these two 
contributions.  
\begin{equation}
n_{0,\pm} = n_{0,\pm}^{(\rm c)} + n_{0,\pm}^{(\rm inc)}, 
\end{equation}
where ``c'' and ``inc'' stands for ``coherent'' and ``incoherent'', 
respectively.  
The variances of the numbers of coherently produced neutral or charged pions 
are enhanced while those of incoherent production are not.  
\begin{equation}
(\Delta n_{0,\pm}^{(\rm c)})^2 = \epsilon_{0,\pm} \langle n_{0,\pm}^{(\rm c)} 
\rangle , \qquad 
(\Delta n_{0,\pm}^{(\rm inc)})^2 = \langle n_{0,\pm}^{(\rm inc)} \rangle,  
\end{equation}
where $\epsilon_{0,\pm}$ are of order $\langle n \rangle \gg 1$ 
(cf.~Eq.~(\ref{e1}), (\ref{e2})).  
Then one can calculate the variance of the sum of the two contributions.  
\begin{equation}
(\Delta n)^2=(\Delta n_{0,\pm}^{(\rm c)})^2+(\Delta n_{0,\pm}^{(\rm inc)})^2
= \tilde\epsilon_{0,\pm} \langle n \rangle, \qquad 
\tilde\epsilon_{0,\pm} = \chi \epsilon + (1-\chi) = 1+ \chi(\epsilon - 1),   
\end{equation}
with 
\begin{equation}
\chi = \langle n_{0,\pm}^{(\rm c)} \rangle / \langle n_{0,\pm} \rangle, \qquad 
1- \chi = \langle n_{0,\pm}^{(\rm inc)} \rangle / \langle n_{0,\pm} \rangle.
\end{equation}
The parameter $\chi$ measures the fraction of pions which are coherently 
produced: $\chi=1$ when all the pions are from D$\chi$C, while $\chi=0$ when 
all of them are independently emitted.  
Obviously, the more incoherent pions in the sample, the smaller is the 
enhancement factor $\tilde\epsilon_{0,\pm}$.   

While these incoherently emitted pions dilute our signatures for D$\chi$C, 
they have different momentum spectra from those from D$\chi$C.  
D$\chi$C pions, being produced from coherent state, carry low $p_T$.  
The typical $p_T$ is of the order of $1/L$, where $L$ is the size of the 
domain from which the pion originates.  
In contrast incoherently emitted pions can carry high $p_T$.  
Therefore applying a low $p_T$ cut can minimize the noise from incoherent pion 
emissions.  

It is also advantageous to measure the rapidity of the pions and count 
their numbers in narrow rapidity windows.  
Bear in mind that the rapidities of the pions are, up to small dispersions, 
equal to that of the original domain.  
As we have mentioned, it is probable that many domains are formed in a 
single collision, and all these domains may have different rapidities. 
For example, the domains at the surface of the ``fire ball'' are moving 
with high speed relative to the domains at the center.  
By binning the pions according to their rapidities, one can partially 
separate the pions from different domains, and the signals are enhanced as 
a result.  

\bigskip

In summary, we suggest the following procedure in looking for signatures 
of D$\chi$Cs.  

$\bullet$ Count the number of neutral or charged pions {\it event by event\/} 
from heavy ion collision experiments and measure their individual transverse 
momenta and rapidities.  

$\bullet$ Apply a low $p_T$ cut to suppress the noise due to uncorrelated 
pion emission. 

$\bullet$ Bin the events in different rapidity windows.  

$\bullet$ In each rapidity window, plot the number of events {\it vs.~}the 
number of neutral or charged pions in histograms.  

$\bullet$ Evaluate the mean, $\langle n_{0,\pm} \rangle$, and the variance, 
$(\Delta n_{0,\pm})^2$, in each rapidity window.  

$\bullet$ If we find $(\Delta n_{0,\pm})^2$ is substantially larger than 
$\langle n_{0,\pm} \rangle$, then we are seeing possible signatures from 
D$\chi$Cs.  

The above procedure allows us to search for signatures from D$\chi$Cs by 
counting only the charged pions.  
This is important as, with our present technology, it is difficult to count 
the number of $\pi_0$'s in a momentum bin, which would mean reconstructing 
all the pions from photons --- a formidable task.  
On the other hand, with great experimental effort, it may be possible to 
count the neutral pions as well in the future.  
In that case, we will be able to distinguish D$\chi$C formation from other 
mechanisms of coherent pion productions.  
For example, one can count $n_t$, the number of pions (both neutral and 
charged) in each rapidity window.  
For D$\chi$C formation, or any other mechanisms of coherent pion productions 
where the field is aligned with a random direction in the four-dimensional 
chiral space $(\pi_x,\pi_y,\pi_0,\sigma)$, the fluctuation of $n_t$ is 
large.  
On the other hand, for mechanisms of coherent pion productions where the 
field is aligned with a random direction in the three-dimensional isospace 
$(\pi_x,\pi_y,\pi_0)$ but without involving the $\sigma$ direction 
({\it i.e.}, the scenario discussed in Ref.~\cite{GGM}), it is 
straightforward to show that the fluctuation of $n_t$ is small.  

There are clearly mechanisms distinct from the D$\chi$C which can enhance 
pion number fluctuations beyond the most naive uncorrelated statistical 
estimate. 
For example, resonances produce more than one pion per decay and hence 
enhance fluctuations.  
Presumably these mechanisms lead to fluctuations which are
characteristically small, with a variance $\sim a/{\cal N}$ where $a$ is 
of order unity (compare with Eq.~(\ref{wide}), where the variance scales 
like $\langle n \rangle /{\cal N}$, where $\langle n \rangle \gg 1$).  
If the pion number fluctuations due to the D$\chi$Cs discussed here are 
small, it will be difficult to distinguish D$\chi$C induced fluctuations from 
fluctuations coming from other processes.  
It is therefore useful to have at least a qualitative estimate of 
background due to these competing processes.  
Some of these backgrounds have been calculated; for example, 
in Ref.~\cite{SRS} for the kinematics of SPS collisions.  
We note, however, that such calculations are necessarily highly detail 
dependent.  
For our purposes, they are useful in that they may give an indication of the 
scale of competing processes.  
To see clear evidence of D$\chi$C production {\it via\/} fluctuations, one 
must be fortunate enough so that the characteristic scale of the D$\chi$C 
fluctuations (in some kinematical window) greatly exceeds fluctuations from 
competing mechanisms.  
Whether or not this is case, cannot be deduced reliably from theory given our
present state of knowledge.  
Ultimately, this must be resolved {\it via\/} experiment.   

In summary, we have constructed signatures for D$\chi$C formation in heavy 
ion collisions which do not require counting the number of neutral pions.  
Instead, we suggest counting the number of charged pions produced, and 
a large fluctuation would be a signal of D$\chi$C formation.  
We believe these new signatures will be useful in searches for D$\chi$C at 
RHIC and LHC.  

\vfill\eject 

\section{Enhancing  Signatures  for Disoriented Chiral Condensates Via 
Conditional Probabilities}

In the previous section we discussed how charged pion fluctuations could be 
used as a possible signature of D$\chi$C formation. 
There remains, however, a serious potential problem in picking signal out of 
the noise if there is a large incoherent pion contribution in even the 
``best'' kinematic windows.   
If it is possible, however, to reconstruct neutral pions reliably, one can 
enhance the signal-to-noise ratio by a judicious choice of data cuts.  
In this section, we discuss how this can be achieved.
We study the distribution of $R$ under the condition that the ``first $k$ 
pions'' in our kinematical window are all neutral, where by the ``first $k$ 
pions'' we mean the $k$ pions with the lowest $p_T$.    
In contrast to the usual result $\langle R \rangle = 1/3$, the conditional 
expectation values are substantially shifted from $1/3$, where the noises due 
to incoherently emitted pions are peaked.  
As a result, one can greatly enhance the signal-to-noise ratio by making cuts 
to retain only those events with the ``first $k$ pions'' are all neutral.  

\subsection{Conditional Probabilities for a Single Domain}

Recall that we have, in Sec.~II, calculated the probability distributions 
$f_0(g_0)$ and $f_\pm(g_\pm)$ for a single domain of D$\chi$C. 
One can also study the probability distribution of the aforementioned 
variable $R = n_0/(n_0+n_\pm) = g_0/(g_0+g_\pm)$ which represents the 
fraction of neutral pions among all pions emitted from the D$\chi$C.  
By reparametrizing distribution (\ref{j}) in terms of $R=\cos^2\theta$ and 
$g_t=g_0+g_\pm=\sin^2\psi$ (the subscript $t$ stands for total), one finds 
the following distributions.  
\begin{equation}
d^2 P(g_t,R) = {1\over \pi} \, \sqrt{g_t\over R(1-g_t)}\;dg_t\,dR, 
\end{equation}
with the marginal distributions 
\begin{eqnarray}
dP(g_t)&= f_t(g_t)\,dg_t=&{2\over\pi}\sqrt{g_t\over 1-g_t}\,dg_t,\\
dP(R)&= f_R(R)d_R = &{1\over 2\sqrt{R}}dR.  
\label{FR}
\end{eqnarray}
Note that while $f_0$ is drastically skewed towards the low end and 
$f_\pm$ is flat, $f_t$ is skewed towards the high end.  
This may sound counter-intuitive, but one must bear in mind that the emissions 
of neutral and charged pions are not independent and there is no contradiction.
Also note that $f_R$ is exactly as predicted in 
Ref.~\cite{And,KK,An1,An2,Bj1,Bj2,RW1,R} (cf.~Eq.~(\ref{PofR})).
We will rename this probability function as $F_0(R)$, where the subscript 
``0'' reminds us that the D$\chi$C is an isosinglet by construction.  
The distribution is plotted in Fig.~2a.  
It is obvious that the shape is qualitatively different from the 
Poisson--Gaussian distribution due to incoherent emissions.  
The expectation value of $R$ is $1/3$,  
\begin{equation}
\langle R \rangle_0 \equiv \int_0^1 R F_0(R) dR = 1/3, 
\label{1/3}
\end{equation}
which has the simple interpretation that it is equally likely for the pion 
to be a $\pi_0$, $\pi_+$ or $\pi_-$, and hence on average a third of the 
pions are neutral.  

This distribution is a consequence of the fact that we have assumed the 
D$\chi$C to be an isosinglet, a reasonable assumption on physical grounds.  
However, let's consider the distribution of $R$ after the D$\chi$C emits a 
single $\pi_0$.  
The density matrix after the emission, which can be written as $\lambda \; 
\pi_0 \, \rho_{D\chi C} \, \pi_0^\dag$, where $\lambda$ is a normalization 
constant, $\rho_{D\chi C}$ is the density matrix defined in Eq.~(\ref{domain}) 
and $\pi_0$ annihilates a neutral pion.  
This new density matrix is not an isosinglet.  
It is straightforward to show that the probability distribution for this 
state is 
\begin{equation}
dP = {\textstyle{1\over2}\sin\theta \cos^2\theta d\theta\over 
\int\textstyle{1\over2}\sin\theta \cos^2\theta d\theta} = 
{\textstyle{1\over2} R \cdot R^{-1/2} dR \over 
\int_0^1 \textstyle{1\over2} R \cdot R^{-1/2} dR}
\end{equation}
or equivalently, 
\begin{equation}
F_1(R) \equiv f(R|\hbox{1st pion is neutral}) 
= R F_0(R) \bigg/\int_0^1 R F_0(R) dR = \textstyle{3\over2} R^{1/2}.
\end{equation}
The distribution $F_1(R)$ is plotted in Fig.~2b, which is 
drastically different from $F_0(R)$.  
The distribution is skewed towards the high end, while $F_0(R)$ is skewed 
towards the low end.  
Moreover, the expectation value of $R$ is clearly pushed up:  
\begin{equation}
\langle R \rangle_1 \equiv \int_0^1 R F_1(R) dR = 3/5.  
\label{3/5}
\end{equation}
So we have arrived at the intriguing conclusion that, if the ``first pion'' 
emitted from a isosinglet D$\chi$C is neutral, 60\% of the pions 
subsequently emitted from the D$\chi$C are neutral, a huge enhancement from 
the original expectation of 33\%.  

\bigskip

\begin{figure}
\epsfig{file=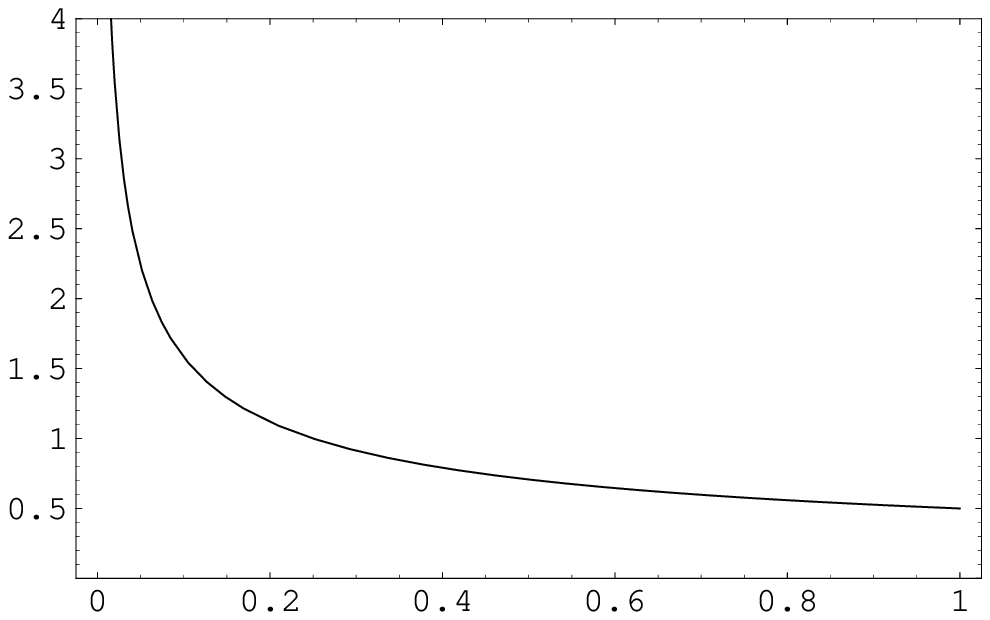, width=3.4in} 
\epsfig{file=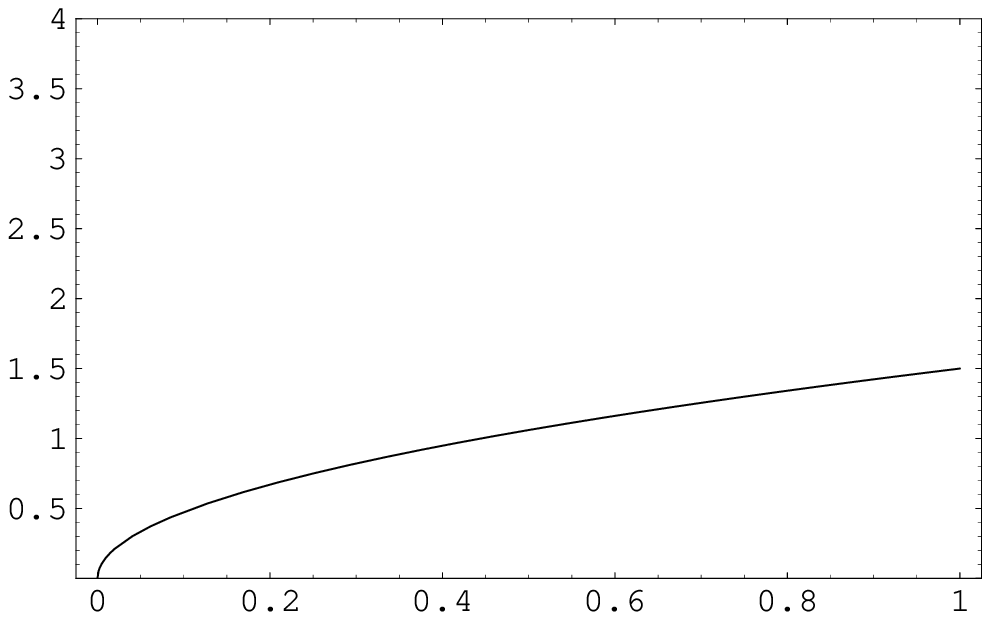, width=3.4in} 

\hskip 115pt Fig.~2a \hskip 215pt Fig.~2b

\bigskip

\epsfig{file=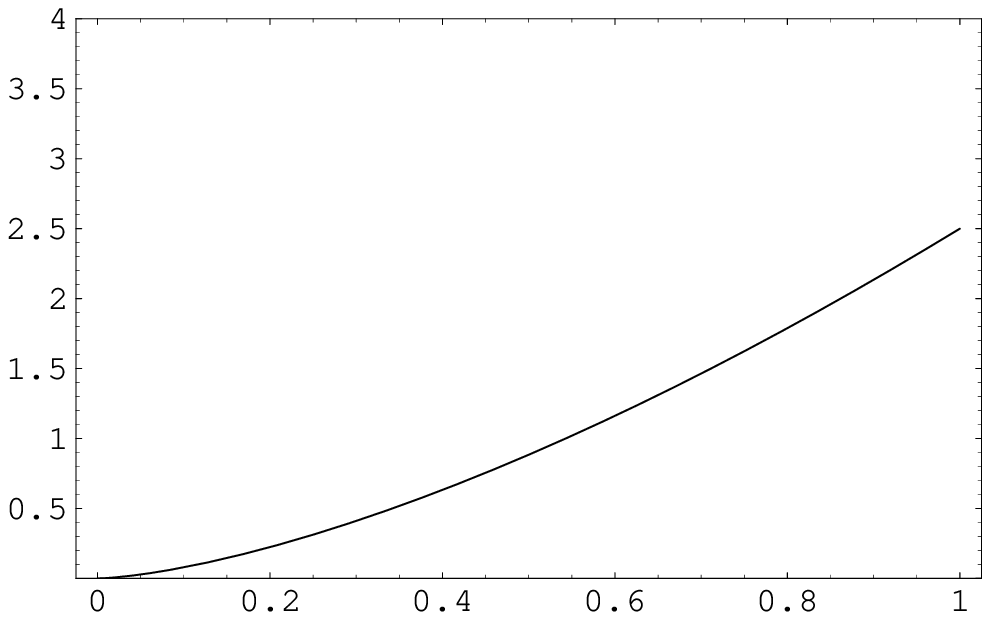, width=3.4in}
\epsfig{file=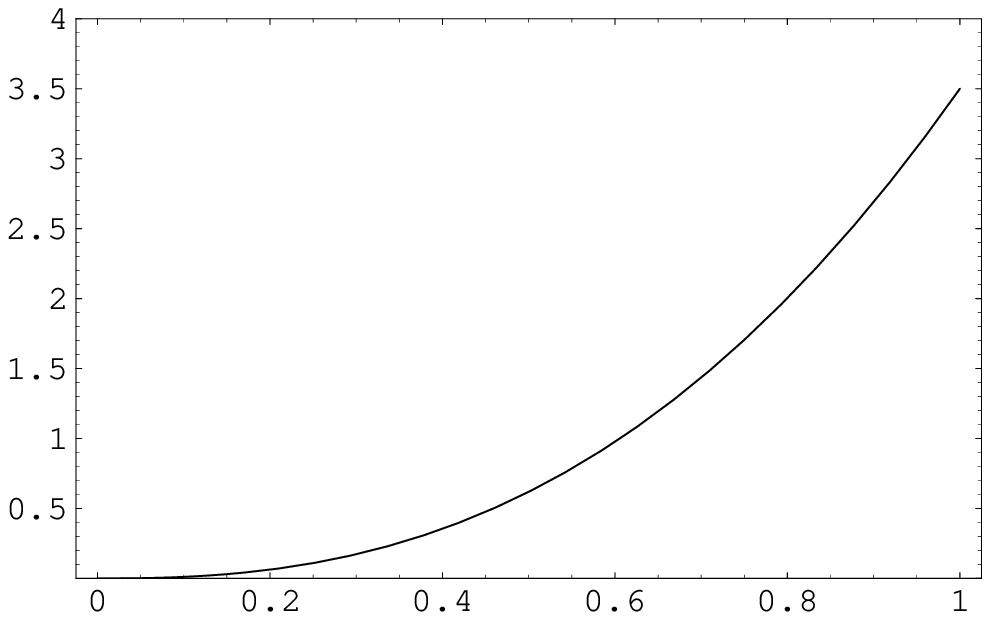, width=3.4in} 

\hskip 115pt Fig.~2c \hskip 215pt Fig.~2d

\bigskip

\caption{The probability distribution functions $F_k(R)$ for different values 
of $k$.
Fig.~2a, b, c and d are $F_0(R)$, $F_1(R)$, $F_2(R)$ and $F_3(R)$, 
respectively.  }
\end{figure}

\bigskip

This extraordinary statement certainly deserves more discussion.  
First, what is our criterion to decide which is the ``first pion''?  
The answer is simple: it can be any criterion.
It does not matter as long as it is {\it a priori\/} equally likely to be a 
$\pi_0$, a $\pi_+$ or a $\pi_-$.  
The derivation just depends on our removing a neutral pion from the isosinglet 
D$\chi$C.  
It can be the first pion emitted in time, or the last one emitted in time, or 
even the 17th emitted in time.  
The criterion can also be unrelated to the order of emission.  
For example, we can choose the ``first pion'' to be the one with the smallest 
polar angle.  
One can use any of these criteria to identify the ``first pion'', and if 
it turns out to be neutral, then the $R$ distribution of subsequent emissions 
is always given by $F_1(R)$, provided we are in the large number limit.  
However, this is only true in this idealized scenario, when all the pions 
are coming from a simple D$\chi$C domain.  
In reality, some of the pions come from incoherent emission, and if the 
``first pion'' turns out to be incoherently emitted, the expectation value 
of $R$ of the remaining pions is still going to be $1/3$, not $3/5$.  
As a result, we want to choose our criterion in such a way that the ``first 
pion'' is likely to originate from the D$\chi$C and not from incoherent 
emissions.  
Since D$\chi$C pions by hypothesis have low $p_T \sim 1/L$, where $L$ is the 
size of the domain, a natural choice is to use the pion with the lowest $p_T$ 
as our ``first pion''.  

After clarifying the meaning of the term ``first pion'', we move on to 
discuss the physical origin of the modification of the probability 
distribution of $R$.  
In a nutshell, we are seeing the physics of (iso)spin alignment due to 
Bose condensation.  
To illustrate the point, let us first consider the following apparently 
unrelated Stern--Gerlach experiment.  
Consider a large number of massive spin-1 particles, which for concreteness 
will be called deuterons.  
Initially they are all polarized along a randomly chosen direction 
$\vec n$, which is {\it a priori\/} equally likely to be any direction 
in three dimensional space.  
In other words, $\vec S \cdot \vec n = 0$ for all the deuterons.  
Now let us pick one of these deuterons and pass it through a Stern--Gerlach 
spectrometer which measures $S_z$, the spin along the $z$-axis.  
What is the probability that the measurement gives $S_z=0$?  
The answer is clearly $1/3$, as the cases for $S_z=+1$, 0 and $-1$ are 
equally likely.  
On the other hand, if the measurement on the first deuteron gives $S_z=0$, 
what is the conditional probability for the next deuteron to pass through 
the Stern--Gerlach spectrometer also to be measured to have $S_z=0$?  
The answer this time is no longer $1/3$.  
The spins of all the deuterons are aligned along the same direction $\vec n$, 
and that the first deuteron is measured to have $S_z=0$ suggests $\vec n$ 
is more probable to be more or less aligned along $\vec z$ than otherwise.  
As a result, the conditional probability is no longer $1/3$, but can be 
easily shown to be $3/5$, which is exactly the predicted value for $\langle 
R \rangle_1$ in Eq.~(\ref{3/5}).  
The situation for a single domain of D$\chi$C is analogous, with 
isospin aligned pions instead of spin aligned deuterons.  
By construction, the pions in a D$\chi$C domain are isospin aligned, and by 
the same analysis, we have shown that the knowledge of the ``first pion'' 
being neutral can dramatically modify the conditional probability 
distribution of $R$.  

One can also consider the conditional probability distribution of $R$ 
in the case that the ``first pion'' is charged.  
Note that
\begin{equation}
F_0(R) = \textstyle{1\over3}\big(f(R|\hbox{1st pion is a $\pi_+$}) + 
f(R|\hbox{1st pion is a $\pi_-$}) + f(R|\hbox{1st pion is a $\pi_0$})\big), 
\end{equation}
and hence, since $F_1(R) = f(R|\hbox{1st pion is a $\pi_0$})$, 
\begin{equation}
\tilde F(R) \equiv f(R|\hbox{1st pion is charged}) = \textstyle{3\over2} 
F_0(R) - \textstyle{1\over2} F_1(R) = \textstyle{3\over4} (1-R) R^{-1/2}.
\end{equation}
The expectation value of $R$, given that the ``first pion'' is charged, can be 
easily shown to be $1/5$.  
As a consistency check, one can calculate $\langle R \rangle$, the expectation 
value of $R$ regardless of the species of the ``first pion''.  
Since the ``first pion'' is twice as likely to be charged as to be neutral, 
\begin{equation}
\langle R \rangle_0 = \textstyle{1\over3}(\textstyle{3\over5}
+2 \times \textstyle{1\over5}) = \textstyle{1\over3}, 
\end{equation}
agreeing with Eq.~(\ref{1/3}).  

Lastly, we will study the conditional probability distribution of $R$ given 
that the $k$ pions with the lowest $p_T$, which will be hereafter referred to 
as the ``first $k$ pions'', are all neutral.  
It is straightforward to show that in this case 
\begin{equation}
dP = {\textstyle{1\over2}\sin\theta \cos^{2k}\theta d\theta\over 
\int\textstyle{1\over2}\sin\theta \cos^{2k}\theta d\theta} = 
{\textstyle{1\over2} R^k \cdot R^{-1/2} dR \over 
\int_0^1 \textstyle{1\over2} R^k \cdot R^{-1/2} dR}. 
\end{equation}
and 
\begin{equation}
F_k(R) \equiv f(R|\hbox{1st $k$ pions are all neutral}) 
= R^k F_0(R) \bigg/\int_0^1 R^k F_0(R) dR = (k+\textstyle{1\over2}) R^{k-1/2}.
\end{equation}
The distributions $F_2(R)$ and $F_3(R)$ are plotted in Fig.~2c and d,  
respectively.  
One can see that as $k$ increases, the distribution is more and more skewed 
towards the high end.  
As a result, the expectation value of $R$ increases with $k$.  
\begin{equation}
\langle R \rangle_k \equiv \int_0^1 R F_k(R) dR = (2k+1)/(2k+3).  
\label{R}
\end{equation}
It is useful to define $Q$ as the ratio of the number of $\pi_+$ 
to the number of total pions emitted.  
By symmetry it is also the ratio of the number of $\pi_-$ to the number of 
total pions emitted, and since $R+2Q=1$, 
\begin{equation} 
\langle Q \rangle_k = 1/(2k+3). 
\label{Q} 
\end{equation}

From the above analysis, the prescription to enhance the collective signal 
is quite clear.  
One should make successive cuts on the data sample on the condition that 
the $k$ pions with the lowest $p_T$ are all neutral, and measure $\langle 
R \rangle_k$ after each cut to see if it increases as predicted in 
Eq.~(\ref{R}).  
This result, however, depends on the assumption that we have only a single 
domain of D$\chi$C without any contamination due to incoherent pion 
emissions.  
Since this assumption is unrealistic for heavy ion collision experiments, 
the scenario we studied in this section is only an idealized situation.  
In the next section, we will discuss more realistic scenarios.  

\subsection{The Effects of Multidomain Formation and Incoherent Emissions}

The scenario considered in the last subsection is highly unrealistic in at 
least two ways.
First, as discussed in the introduction, single domain D$\chi$C formation is 
highly unlikely.  
For a realistic treatment one must study D$\chi$C formation with more 
than one domain, each pointing in a different direction in the isospace.  
Moreover, we have neglected the effect of incoherently emitted pions, which 
have very important effects.  
If the neutral ``first pion'' is incoherently emitted, the $R$ distribution 
of the remaining pions is described by $f_0(R)$, instead of $f_1(R)$ when 
the ``first pion'' comes from the D$\chi$C.  
In this section, we will incorporate these two effects and see how the 
predictions above are modified.  

We will study the expectation value of $R$, or equivalently the expectation of 
$Q$, for a situation described by the following parameters.  
The coherent fraction $\chi$ is the fraction of pions which originate from 
D$\chi$C domains, so that when $\chi=1$, all pions are coherently emitted, 
and when $\chi=0$, all pions are incoherently emitted.  
We will consider the case where there are $N$ domains, all containing an equal 
number of pions\footnote
{This assumption of all domains having the same number of pions is unrealistic 
but is made for illustrative purposes.  
The effects of unequal domain sizes will be briefly discussed below.}, which 
will be assumed to be large.    
Each domain is described by an isosinglet density matrix, but the isospins of 
pions in different domains are uncorrelated.    
Now the question is: if the ``first $k$ pions'' in this channel are all 
neutral, what are the expectation values of $R$ and $Q$ among the rest of the 
pions?  

The answer turns out to be the following expression:   
\begin{mathletters}
\begin{equation}
\langle R \rangle = \textstyle{1\over3} + 2\Delta, \qquad 
\langle Q \rangle = \textstyle{1\over3} - \Delta.
\end{equation}
The shift $\Delta$ is given by 
\begin{equation}
\Delta =  \chi \sum_{j=0}^k P_j \,  ({1\over 3} - {1\over 2j+3}) 
= \chi \big({1\over 3} - \sum_{j=0}^k P_j \,  {1\over 2j+3} \big),    
\end{equation}
where 
\begin{equation}
P_j = {k \choose j} \; p^j \, (1-p)^{k-j}, \qquad p=\chi/N. 
\end{equation}
\label{result}
\end{mathletters}
Each term in this formula has a simple interpretation: 

$\bullet$ The expectation value $\langle R \rangle$ is always $1/3$ for the 
incoherently emitted pions.  
Only the pions coming from the domains are affected by isospin alignment; 
hence the outstanding factor of $\chi$.  

$\bullet$ Each coherently emitted pion comes from one of the domains, which 
will be called domain X.  
How many of the ``first $k$ pions'' also come from domain X?  
The probability for each pion coming from domain X is $p=\chi/N$, and the 
probability that $j$ of the ``first $k$ pions'' coming from domain X is 
$P_j = {k \choose j} \, p^j \, (1-p)^{k-j}$.  

$\bullet$ Given that $j$ of the ``first $k$ pions'' is coming from domain X, 
the conditional expectation value of $Q$ decreases from $1/3$ to $1/(2j+3)$, 
while the conditional expectation value of $R$ increases by twice the above 
quantity.   

In passing, we note that $\Delta$ can also be expressed as an integral or 
the hypergeometric function $_2F_1$: 
\begin{eqnarray}
\Delta &=& \chi \Big({1\over 3} - {1\over \sqrt{p^3}} 
\int_0^{\sqrt{p}} dz \; z^2 (z^2 + 1 - p)^k \Big) \nonumber \\  
&=& \chi \Big({1\over 3} - {\cos^{2k+3}\Theta \over \sin^3\Theta} 
\int_0^\Theta d\vartheta\; {\sin^2\vartheta \over \cos^{2k+4}\vartheta} \Big), 
\qquad \tan^2\Theta = {\chi/N \over 1- \,\chi/N} \nonumber \\
&=& {1\over 3} \chi \Big(1-(1-p)^k \, 
_2F_1[{3\over2},-k,{5\over2},{-p\over 1-p}] \Big).  
\end{eqnarray}
In Fig.~3, we have made contour plots of $\Delta=1/30$ (such that $\langle R 
\rangle = 0.4$ and $\langle Q \rangle = 0.3$), for $k=1,\dots,5$ in the 
$(\chi,1/N)$ parameter space.  
The horizontal axis is the coherent fraction $\chi$ while the vertical axis is 
$1/N$ where $N$ is the number of domains.  
Both $\chi$ and $1/N$ range from 0 to 1.  
Thus, for example, with 3 domains and $\chi=0.6$, in order to have $\Delta 
\geq 1/30$ we must have $k\geq 3$.  

Equations (\ref{result}) illustrate the main results of this section.  
One can see that, without any D$\chi$C formation, $\chi=0$ (corresponding to 
the left edge of Fig.~3), $\Delta$ vanishes, and $\langle R \rangle = 
\langle Q \rangle = 1/3$ as expected.  
The bottom edge of the plot corresponds to $N\to\infty$ and also gives 
$\Delta = 0$ for any finite value of $k$.  
The shift $\Delta$ is largest for a single domain of D$\chi$C without 
any noise due to incoherently emitted pions, {\it i.e.}, when $\chi=N=1$ (the 
top right corner of the contour plots), giving $\Delta = 1/3 \, - \, 
1/(2k+3)$ and reproducing Eqs.~(\ref{R}) and (\ref{Q}).  
For fixed values of $(\chi,N)$, $\Delta$ increases with $k$, accounting for 
the spreading of the parameter space with $\Delta > 0.4$ as $k$ increases  
from 1 to 5 in Fig.~3.

\bigskip

\begin{figure}

\centerline{ \epsfig{file=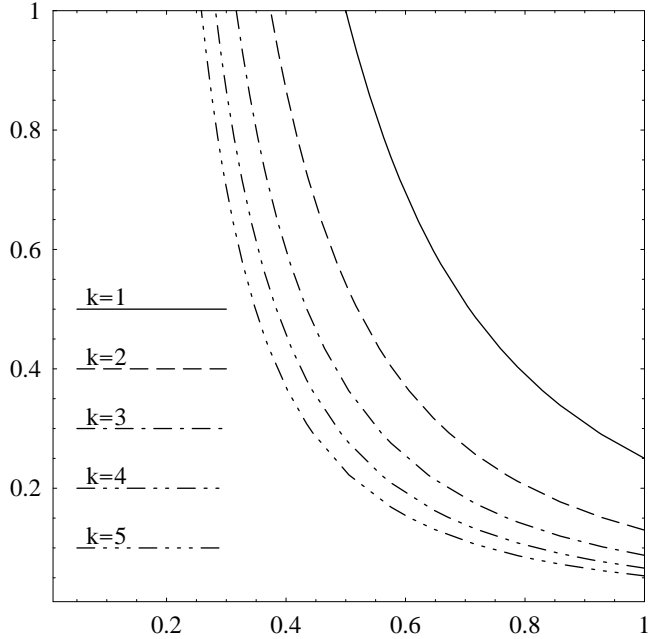, width=3.4in} }

\bigskip

\caption{Contour plots of $\Delta = 1/30$ (such that $\langle R \rangle =
0.4$ and $\langle Q \rangle = 0.3$) for different values of $k$.  
The horizontal axis is the coherent fraction $\chi$, while the vertical axis 
is $1/N$ where $N$ is the number of domains.  
Both $\chi$ and $1/N$ range from 0 to 1. 
The curves, from top right to bottom left are for $k=1$, 2, 3 4 and 5,  
respectively. 
The shift $\Delta$ is larger than $0.4$ above the curves and smaller than 
$0.4$ below the curves. }
\end{figure}

\bigskip

One expects that when the number of D$\chi$C domains is large ($N\gg1$) or 
when most of the pions are incoherently emitted ($\chi\ll 1$), it will be 
difficult to observe clear signals of D$\chi$C formation.  
However, in such situations $\chi/N$ is small and $\Delta$ is dominated by 
the $j=1$ term (the $j=0$ term always identically vanishes) and 
\begin{equation}
\Delta = {2 \chi^2 k \over 15 N} + {\cal O}({\chi^3\over N^2}).  
\label{small}
\end{equation}
Thus a large $k$ may make up for a small coherent fraction $\chi$, or a large 
number of domains $N$, and enhance $\Delta$, which describes the shift of 
$\langle R \rangle$ and $\langle Q \rangle$ from $1/3$, to an experimentally 
measurable magnitude.  
From the form of Eq.~(\ref{small}), one expects this shift to be substantial 
whenever $k \sim N/\chi^2$. 
However, even for a value of $k$ as small as $N/4\chi^2$, $\Delta = 1/30 + 
{\cal O} (\chi^3/N^2)$, which translates to $\langle R \rangle = 0.4$ and 
$\langle Q \rangle = 0.3$ --- a substantial deviation from the incoherent 
case.  
This suggests one should make successive cuts for events where the $k$ pions 
with lowest $p_T$ are all neutral, and study $\langle R \rangle$ after each 
cut.  
An increase of $\langle R \rangle$ with $k$ would suggest that D$\chi$C 
domains are formed.  

Equation (\ref{small}) appears to suggest that one can increase $\Delta$ to 
an arbitrarily large magnitude by choosing a sufficiently large value of $k$. 
Of course this is not true.  
Equation (\ref{small}) is obtained as the leading term in a $\chi/N$ 
expansion, but when $k \to \infty$, this expansion breaks down as terms of 
higher order in $\chi/N$ are enhanced by factors of $k \choose j$.  
We can easily see that 
\begin{equation}
\Delta \to \textstyle{1\over3}\chi, \quad 
\langle R \rangle \to \textstyle{1\over3}+\textstyle{2\over3}\chi, \quad
\langle Q \rangle \to \textstyle{1\over3}(1-\chi), \qquad k\to\infty 
\hbox{ with $\chi$ and $N$ fixed.} 
\label{limit}
\end{equation}
In other words, our signal enhancement scheme is fundamentally limited by the 
amount of noise due to incoherently emitted pions.  
When $\chi$ is small, most of the pions are incoherently emitted, and for 
them, $\langle R \rangle$ is always around $1/3$ regardless of what cuts one 
makes.  
On the other hand, the large $k$ limit of $\langle R \rangle$ does not depend 
on $N$, the number of D$\chi$C domains.  
Recall that we have several distinctive signatures, like the $R$ distribution 
in Eq.~(\ref{PofR}) and the conditional expectation values for $R$ described 
in the previous section, for a single domain of D$\chi$C, where all the pions 
in the D$\chi$C are isospin aligned.  
With multi-domain formation, where the pions in different domains may point 
to different directions in isospace, the effect of isospin alignment is 
greatly washed out.  
However, given that the ``first $k$ pions'' are all neutral with $k\gg N$, 
it is probabilistically extremely likely that each of the $N$ domains is 
the origin of some of these ``first $k$ pions''.  
As a result, each of these $N$ domains are well-aligned along the $\pi_0$ 
direction, and hence also well-aligned with each other.  
As $k\to\infty$, the $N$ domains look more and more like a single big 
domain in the $\pi_0$ direction, which is the case where the signature is 
the most dramatic.  
In short, with large $k$, our cuts are picking out the events where the 
signals are the strongest, and hence resulting in a large signal-to-noise 
ratio.  

In above we have assumed the sizes of all $N$ domains are identical for 
illustrative purposes.  
A more realistic treatment would have $N$ domains, each with different 
sizes $p_i$, $1\leq i \leq N$, such that $\sum_i p_i = \chi$.  
(The size of a domain is defined to be the fraction of pions which 
originate from this particular domain.)  
Then Eqs.~(\ref{result}) are generalized to 
\begin{equation}
\Delta = {1\over 3} \chi - \sum_{i=1}^N p_i \; \sum_{j=0}^k {k \choose j} 
p_i^k (1-p)^{k-j} \; {1\over 2j+3}.  
\end{equation}
In the weak signal limit, {\it i.e.}, when all the $p_i$'s are small, 
Eq.~(\ref{small}) gets modified to 
\begin{equation}
\Delta = {2k \over 15} \sum_{i=1}^N p_i^2 + {\cal O}({\chi^3\over N^2}) 
= {2k \over 15} \overline p + {\cal O}({\chi^3\over N^2}) , 
\end{equation}
where $\overline p = \sum_i p_i^2$ has the following nice interpretation: 
$\overline p$ is the average over all pions (both coherently and incoherently 
emitted) of the sizes of the originating domains, which is $p_i$ for a pion 
from domain $i$ and zero for an incoherently emitted pion.   
For $N$ domains of equal sizes, $\overline p = \chi^2/N$ and Eq.~(\ref{small}) 
is recovered.  
Again we see that $\Delta$ grows linearly with $k$ in the weak signal limit.  
As $k\to\infty$ the shift $\Delta$ is again limited by the bound 
(\ref{limit}), which applies also for the cases of unequal domain sizes.  

\bigskip

To recapitulate, we suggest the following experimental procedures:  

$\bullet$ Count the number of neutral and charged pions {\it event by event\/} 
from heavy ion collision experiments and measure their individual transverse 
momenta and rapidities.  

$\bullet$ Apply a low $p_T$ cut to suppress the noise due to uncorrelated 
pion emission. 

$\bullet$ Bin the events in different rapidity windows.  

$\bullet$ In each rapidity window, calculate the expectation value 
$\langle R \rangle$.  

$\bullet$ Make a cut to retain only events where the pion with the lowest 
$p_T$ is neutral.  

$\bullet$ Calculate, in each rapidity window, the expectation value 
$\langle R \rangle$ for all remaining pions in all events which survive the 
cut.  

$\bullet$ Make another cut on the surviving events to retain only those where 
the pion with the second lowest $p_T$ is also neutral.  

$\bullet$ Again, calculate in each rapidity window the expectation value 
$\langle R \rangle$ for all remaining pions in all events which survive the 
cuts. 

$\bullet$ Repeat the above prescription of making successive cuts to retain 
only events in which the pion with the next lowest $p_T$ is also neutral, 
and calculate $\langle R \rangle$ for each rapidity window after each cut.  
If we find $\langle R \rangle$ deviates from $1/3$ then we are seeing 
signatures from D$\chi$Cs.  

Note that this prescription requires reconstructions of $p_T$'s of 
individual pions, both charged and neutral.  
We have also presumed that the coherent fraction $\chi$ and the number of 
domains formed $N$ are roughly the same for each event.  
(More specifically, the probability distributions for $\chi$ and $N$ are 
narrow peaked.)  

By applying these successive cuts, we are retaining the events with D$\chi$C 
formation {\it and\/} most of the pions are well-aligned along the $\pi_0$ 
direction.  
What is being cut are the events with D$\chi$C formation but most of 
the pions are well-aligned along the $\pi_x$ or $\pi_y$ directions, and the 
events where there are incoherent pions with very low $p_T$, which is the main 
source of noise to our signal.  
As a result, these successive cuts are substantially improving the 
signal-to-noise ratio, making it easier to observe D$\chi$C formation.  
On the other hand, just like any other cuts on data to suppress the noises, 
we are giving up on statistics.  
Moreover, for large $k$ we are cutting on rare events so the loss in 
statistics can be severe.  
For the cases where the signal is weak (small coherent fraction $\chi\ll 1$ or 
large number of domains $N\gg 1$) on each cut we are losing about two-thirds 
of the events.  

In conclusion, we have devised new cuts to enhance the signal in searches for 
D$\chi$C.  
These cuts retain only events where the $k$ pions with lowest $p_T$ are 
all neutral.  
We have shown that, after these cuts, the fraction of neutral pions within 
the remaining sample is substantially larger if D$\chi$Cs are formed in the 
heavy ion collision.  

\section{summary}

In this paper, we have discussed two methods for detection of D$\chi$C in 
heavy ion collisions.  
Both methods make use of the fact that all the pions in a D$\chi$C are 
aligned in the O(4) chiral space, and hence also in the O(3) isospace.  
This leads to a larger fluctuation in the number of charged pions produced 
than the case where all the pions are independently emitted.  
As a result, one can look for signatures for D$\chi$C formation from charged 
pion data only.  
The alignment in isospace also suggests that, if the ``first $k$ pions'' are
all neutral, the remaining pions are also much more likely to be neutral 
than otherwise.  
By making cuts accordingly, one can greatly enhance the signal-to-noise ratio. 
We expect these methods will be useful in future searches for signatures of 
D$\chi$C formation in RHIC and LHC.  

\bigskip

Support of this research by the U.S.~Department of Energy under grant 
DE-FG02-93ER-40762 is gratefully acknowledged.

\end{document}